\documentclass{superfri}
\usepackage[hidelinks]{hyperref}

\usepackage{amsmath,amsfonts,amssymb,amsthm}

\newcommand{\bP}{{\bf P}}
\newcommand{\bF}{{\bf F}}
\newcommand{\bU}{{\bf U}}

\begin{document}

\author{Agnieszka Janiuk\footnote{\label{susu} Center for Theoretical Physics PAS, 
Warsaw, Poland}, Konstantinos Sapountzis$^{\ref{susu}}$, J\'er\'emy Mortier$^{\ref{susu}}$ \and Ireneusz Janiuk\footnote{\label{msu} Verifone Sp. Z.o.o., 
Warsaw, Poland}}

\title{Numerical Simulations of Black Hole Accretion Flows}

\maketitle{}

\begin{abstract}
We model the structure and evolution of black hole accretion disks using numerical simulations. The numerics is governed by the equations of general relativistic magneto-hydrodynamics (GRMHD).
Accretion disks and outflows can be found at the base of very energetic ultra-relativistic jets produced by cosmic explosions, so called gamma-ray bursts (GRBs). Another type of phenomena are blazars, with jets emitted from the centers of galaxies.
 
Long-lasting, detailed computations are essential to determine the
physics of these explosions, and confront the theory with potential observables.
From the point of view of numerical methods and techniques, three ingredients need to be considered. First, the numerical scheme must work 
in a conservative manner, which is achieved by solving a set of non-linear equations to advance the conserved quantities from one time step to the next. Second, the efficiency of computations depends on the code parallelization methods. Third, the analysis of results is possible via the post-processing of computed physical quantities, and visualization of the flow properties. This is done via implementing packages and libraries that are standardized in the field of computational astrophysics and supported by community developers.

In this paper, we discuss the physics of the cosmic sources. We also describe our numerical framework and some technical issues, in the context of the GRMHD code which we develop.
We also present a suite of performance tests, done on the High-Performance Computer cluster (HPC) in the Center for Mathematical Modeling of the Warsaw University.

\end{abstract}

\section*{Introduction}
\label{sec:intro}

Astrophysical black holes are ubiquitous in the Universe.
They occupy centers of galaxies \cite{rees1984black}, they may be found in the binary systems
with ordinary stars, where the streams of plasma lead to the phenomenon called a 'microquasar' \cite{mirabel1999sources}, and, finally, they may be responsible for the
most extreme cosmic explosions - the gamma ray bursts \cite{meszaros2006gamma}.
In all these types of sources one common physical process is in work: accretion of matter onto a black hole. It is the most efficient of the known energy release mechanisms, which is orders of magnitude stronger than the nuclear fusion reactions that fuel ordinary stars. The gravitational potential energy in the field of the compact star is governed by its mass-to-radius ratio. Hence, per unit rest-mass energy of the gas fallen into the black hole, we can extract up to almost sixty per-cent of power available to be released in the form of the electromagnetic radiation \cite{king_book}.

Gamma ray bursts (GRBs) are electromagnetic transients observed from the most distant parts of the Universe. Their brightness detected by the human-made telescopes implies that the intrinsic power of the events is enormously large.
During the collapse of a massive star into a black hole in a hyper-nova process a long duration burst ($t\sim 100-1000$ seconds) can be observed. It is required that the progenitor star has enough angular momentum to form an accretion disk around a black hole. Short GRBs ($t \sim 0.01-2$ seconds) are associated with the coalescence of binary neutron stars, which form a black hole as a product of their merger. The transient jets of plasma are generated by central engine, which is composed of a newly formed black hole, surrounded by a remnant disk. These jets are ultimately responsible for the electromagnetic gamma-ray emission, observed by our telescopes.

Another type of sources visible as the ultra-relativistic jets of plasma that emit very high energy radiation spectra, are called blazars. These sources are persistent in nature, and they do not originate in violent explosions.
However, the mechanism of extracting energy from the rotating black hole is the same, and requires the magnetic fields to mediate the process.
The famous Blandford--Znajek mechanism can work as a kind of cosmic 'battery' 
\cite{blandford1977} and requires that the accretion disk is supplied with a 
strong poloidal magnetic field. When the jet is pointing in the direction of 
the Earth, the observer detects phenomenon called a 'blazar', where the highest energy flux is detected due to the boosting effect and collimation of the 
stream in a narrow cone around the line of sight \cite{urry1995unified}. Gamma-ray emission 
of blazars exhibits often a short-timescale variability that lasts from hours 
to days \cite{abdo2010gamma}. This means that the gamma-ray emission from the 
jet is not uniform and short time-scale variability suggest that it occurs 
close to the black hole. This effect is quite similar again to the GRB 
variability, albeit  the timescales are now on the order of a millisecond.

Our work focuses on modeling the structure and evolution of the accretion disk at the base of the jet engine, which is composed of a highly magnetized plasma accreting onto a black hole.
In order to construct a physical model of such disk, we need to solve the GRMHD equations.
They are further supplemented by the equation of state (EOS) of the matter.
Its form depends on the particular phenomenon and astrophysical scenario considered.
In the quiescent centers of galaxies, and also in the persistent jet sources, such as blazars, the accreting matter is quite hot, but rarefied, and to a good approximation it can be described with an ideal gas EOS. 
In the GRBs explosive events, the EOS is more complex. Under the conditions of extremely high densities and temperatures, the nuclear reactions have to be taken into account. Furthermore, the matter is highly degenerated, and relativistic hot particles cannot form an ideal gas. The density and temperature are tied to the pressure and internal energy 
in a non-linear way.
Finally, the nuclear processes may occur on the rates faster than the timescale required to establish the statistical equilibrium conditions in the gas. 

All these physical complexities: magnetic fields, general relativity, nuclear reactions, pose a challenge to any kind of numerical scheme.
Different codes have been proposed to cover both the microphysics of the fluids, governed by EOS, and the evolution of magnetized gas in 
the black hole gravitational potential, governed by the GRMHD equations.

The article is organized as follows. In Section \ref{sec:grmhd} we define the physical equations thet are solved by our GRMHD code and we describe the conversion scheme. In Section \ref{sec:kostas} we present the tools used for visualisation of the simulation results. Section \ref{sec:axis} is devoted to the problem of boundary conditions in MHD simulations. In Section \ref{sec:mpi} we discuss the parallelisation methods and compare the performance of the code when different techniques are used. Section \ref{sec:simulations} presents exemplary results of our simulations of a black hole, torus and jet system evolved with the 
GRMHD code, and the results of post-processing simulation with a nuclear reaction network code.
In Section \ref{sec:concl} we discuss our computations in the broader context of recent astronomical discoveries.

\section{GRMHD Equations}
\label{sec:grmhd}

From the steady state based on the analytical, equilibrium solution driven by the main physical parameters of the black hole accretion disk (namely, BH mass, its spin, and size and the mean accretion rate in the torus), as well as the seed configuration of magnetic field, we follow the dynamical evolution. This is achieved by solving numerically the continuity Eq. (\ref{p-number}), the four-momentum-energy conservation Eq. (\ref{four-energy}), and induction Eq. (\ref{induction}) equations in GR framework:

\begin{equation}
\label{p-number}
\frac{1}{\sqrt{-g}} \partial _{\mu} (\sqrt{-g}\ \rho u^{\mu}) = 0 ,
\end{equation}
\begin{equation}
\label{four-energy}
\partial _t (\sqrt{-g}\ T^t_{\ \nu}) = -\partial_i (\sqrt{-g}\ T^i_{\ \nu}) + \sqrt{-g}\ T^{\kappa}_{\ \lambda} \Gamma^{\lambda}_{\ \nu \kappa} ,
\end{equation}
\vspace*{0.1cm}
\begin{equation}
\label{induction}
\partial_t (\sqrt{-g}\ B^i) = -\partial_j (\sqrt{-g}\ (b^j u^i - b^i u^j)) . 
\end{equation}

Here the stress tensor separates into gas and electromagnetic parts:

\begin{equation}
\begin{split}
\hspace{1cm}
T^{\mu\nu} = T^{\mu\nu}_{gas}+T^{\mu\nu}_{EM} \hspace{2.5cm} , \\
T^{\mu\nu}_{gas} = \rho h u^{\mu} u^{\nu}+pg^{\mu\nu} = ( \rho + u + p)u^{\mu} u^{\nu}+pg^{\mu\nu} , \\
T^{\mu\nu}_{EM} = b^{2} u^{\mu} u^{\nu}+\frac{1}{2} b^{2} g^{\mu\nu}- b^{\mu} b^{\nu};  b^{\mu}=u^{*}_{\nu}F^{\mu\nu}  . \hspace{0.5cm}
\end{split}
\label{stress}
\end{equation}

Other symbols in Eq. (\ref{stress}) have their usual meaning:
$u^{\mu}$ is the four-velocity of gas, $u$ is internal energy, $\rho$ is density, $p$ denotes pressure, and  $b^{\mu}$ is the magnetic four-vector. Note that 
in Eq. (\ref{induction}) $B^i$ is the magnetic field three-vector, $b^i$ is the spatial part of the magnetic field four-vector and $u^i$ is the spatial part of the four-velocity.
Finally, $F$ is the Faraday tensor, and in the force-free approximation 
$E_{\nu}=  u^{\nu}F^{\mu\nu} = 0$. 
The space-time metric $g_{\mu \nu}$ is described
in Eq. (\ref{p-number}) with determinant $g\equiv$ Det$(g_{\mu \nu})$ 
and $\Gamma^{\lambda}_{\ \nu \kappa}$ is the spatial connection. 

\subsection{Our Code for the High Accuracy Relativistic MHD}
\begin{figure} 
 	\centering
\includegraphics[width=3.0in]{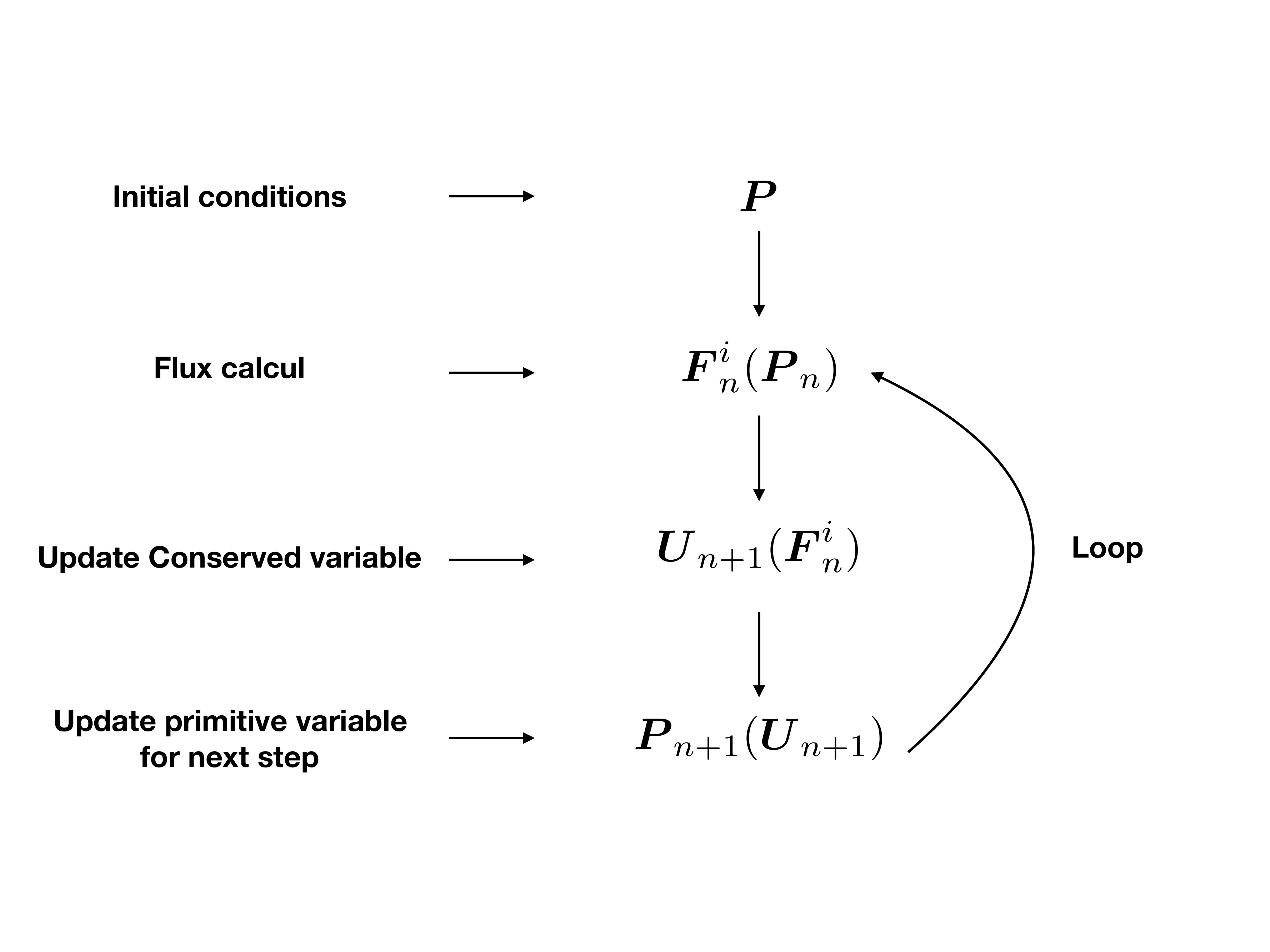}
\caption{Time step of a conservative scheme of order 1 in time. $n$ denotes the time step}
\label{time_step}
\end{figure}

HARM (High Accuracy Relativistic Magneto-hydrodynamics) is a conservative shock capturing scheme, for evolving the equations of GRMHD, developed by C. Gammie et al. \cite{gammie2003harm}. The integrated equations are of the form:

\vspace{0.2cm}
\begin{equation}
\partial_t \boldsymbol{U} (\boldsymbol{P}) = - \partial_i \boldsymbol{F}^i(\boldsymbol{P}) + \boldsymbol{S}(\boldsymbol{P}) ,
\label{eq:utoprim}
\end{equation} 
\vspace{0.2cm}

where $\boldsymbol{U}$ is a vector of ``conserved variables'', such as particle number density, or energy or momentum, $\boldsymbol{F}^i$ are the fluxes in finite control volume, and $\boldsymbol{S}$ is a vector of source terms. $\boldsymbol{U}$ is conserved in the sense that, if $\boldsymbol{S} = 0$, it depends only on fluxes at the boundaries. The vector $\boldsymbol{P}$ is composed of ``primitive'' variables, such as rest-mass density, internal energy density, velocity components, and magnetic field components, which are interpolated to model the flow within zones. $\boldsymbol{U}$ and $\boldsymbol{F}^i$ depend on $\boldsymbol{P}$. Conservative numerical schemes advance $\boldsymbol{U}$, then, depending on the order of the scheme, calculate $\boldsymbol{P} (\boldsymbol{U})$ once or twice per time step, as shown in Fig. \ref{time_step}.

Our version of the code works in 2D and 3D. It is fully parallelized using the Message Passing Interface (MPI) library (see Section \ref{sec:mpi} for parellelisation and performance test results) and the output of the simulation is dumped both in ASCII and Hierarchical Data Format (see Section \ref{sec:kostas}).

\subsection{Equation of State}

The equation of state in the plasma is based on equilibrium of the 
nuclear reactions, which, when reached, defines the proton-to-baryon density ratio, and hence the pressure, internal energy and entropy of the gas.

In the hyper-accreting matter in the GRB central engine, temperatures are above $10^{9}-10^{10}$ K, and the plasma is fully ionized, and composed of free nuclei, $n$, $p$, and electron-positron pairs, $e^{+}$, $e^{-}$.
The chemical and pressure balance required by nuclear reactions between these species, namely the electron and positron capture on nuclei, and the $\beta$-decay.
Reactions are in the form:
\begin{eqnarray}
\label{eq:urca}
p + e^{-} \to n + \nu_{\rm e} \nonumber , \\
p + \bar\nu_{\rm e} \to n + e^{+}  \nonumber , \\
p + e^{-} + \bar\nu_{e} \to n \nonumber , \\
n + e^{+} \to p + \bar\nu_{\rm e} \nonumber , \\
n \to p + e^{-} + \bar\nu_{\rm e} \nonumber , \\
n + \nu_{\rm e} \to p + e^{-} \nonumber .
\end{eqnarray}

The rates for these reactions are given by appropriate integrals \cite{reddy1998neutrino}.

The electron neutrinos and anti-neutrinos are the source for cooling of the plasma, in addition to advective and radiative cooling (the latter is in fact negligible, due to the huge opacities for the photons).
The above nuclear processes have been studied in numerous works, devoted to the neutron star Equation of State, and later on in the context of GRB central engines \cite{popham1999hyperaccreting, di2002neutrino, kohri2002can, chen2007neutrino, lee2006accretion, janiuk2004evolution, janiuk2007instabilities, janiuk2013accretion}. 


Other neutrino emission processes that occur at lower rates are: electron-positron pair annihilation, bremsstrahlung, and plasmon decay:
\begin{eqnarray} 
e^{-}+e^{+}\to \nu_{\rm i}+\bar\nu_{\rm i} \nonumber , \\
n+n \to n+n+\nu_{\rm i}+\bar\nu_{\rm i} \nonumber , \\
\tilde \gamma \to \nu_{\rm e}+\bar\nu_{\rm e} \nonumber .
\end{eqnarray}

We calculate their rates numerically, with proper integrals over the distribution function of relativistic, partially degenerated species.
These processes lead to formation of heavy lepton neutrinos, $\nu_{\tau}$ and $\nu_{\mu}$.

\subsubsection{Pressure Components}

In the EOS, contribution to the pressure is by the free nuclei and $e^{+}-e^{-}$ pairs, 
helium, radiation and the trapped 
neutrinos: 
\begin{equation}
P = P_{\rm nucl}+P_{\rm He}+P_{\rm rad}+P_{\nu} ,
\nonumber
\end{equation} where
$P_{\rm nucl}$ includes free neutrons, protons, 
and the electron-positrons:
\begin{equation}
P_{\rm nucl}=P_{\rm e-}+P_{\rm e+}+P_{\rm n}+P_{\rm p} ,
\nonumber
\end{equation}
with
\begin{equation}
P_{\rm i} = {2 \sqrt{2}\over 3\pi^{2}}
{(m_{i}c^{2})^{4} \over (\hbar c)^{3}}\beta_{i}^{5/2}
\left[F_{3/2}(\eta_{\rm i},\beta_{\rm i})+{1\over 2} \beta_{\rm i}F_{5/2}(\eta_{\rm i},\beta_{\rm i})\right] .
\label{eq:pi}
\end{equation}
Here in Eq. (\ref{eq:pi}), $F_{\rm k}$ are the Fermi--Dirac integrals of the order $k$, and
$\eta_{\rm e}$, $\eta_{\rm p}$ and $\eta_{\rm n}$ are the reduced chemical
potentials, $\eta_i = \mu_i/kT$,  
 is the degeneracy parameter (where $\mu_i$ is the standard chemical
potential) 
Reduced chemical potential of positrons is
$\eta_{\rm e+}=-\eta_{\rm e}-2/\beta_{\rm e}$.
Relativity parameters are defined as $\beta_{\rm i}=kT/m_{\rm i}c^{2}$.
EOS computed numerically by solving the balance of nuclear reactions
\cite{yuan2005, janiuk2007instabilities, janiukyuan2010}.

To sum up, the proper description of the hyper-accretion in GRBs 
requires detailed treatment of microphysics. 
Based on the solutions for 
degenerate Fermi gas EOS, with $P(\rho,T)$ non-linear dependence, a
non-trivial transformation between conserved variables and
'primitives' in HARM  due to GRMHD scheme.

The interpolation over the tables of EOS is done (using the Akima-spline method \cite{akima}) during the dynamical simulation
at each and every time step. In order to save the computer power, we usually compute a small matrix of $4 \times 4=16$ points at each grid cell, whenever the value must be interpolated, and only then we store the table. To perform the interpolations with maximum efficiency, we use the multi-threading feature
of the Linux operating system (with \textit{pthread} command).

\subsubsection{Neutrino Cooling}

The effect important for the state of accretion disk is neutrino cooling. The neutrinos of three flavors are emitted via the above weak interactions, and
the neutrino cooling rate is finally given by the two-stream approximation, and includes
the scattering and absorptive optical depths for neutrinos of all three
flavors:
\begin{equation}
Q^{-}_{\nu} = { {7 \over 8} \sigma T^{4} \over 
{3 \over 4}} \sum_{i=e,\mu,\tau} { 1 \over {\tau_{\rm a, \nu_{i}} + \tau_{\rm s} \over 2} 
+ {1 \over \sqrt 3} + 
{1 \over 3\tau_{\rm a, \nu_{i}}}} \times {1 \over H}\; ~[{\rm erg ~s^{-1} ~cm^{-3}]} ,
\nonumber
\end{equation}
 as given by \cite{di2002neutrino}.
This expression assumes that neutrinos are thermalized.
Ideally, neutrino transport should be accounted for
(see e.g. \cite{perego2014}).

\subsection{Conversion Scheme}

The composition-dependent EOS is in our simulations coupled to the conservative scheme.
The HARM (high-accuracy relativistic magneto-hydrodynamics) scheme solves 
equations Eq. \ref{eq:utoprim}
where $\bU$ is a  vector of ``conserved'' variables,
(i.e., the number density, energy or momentum density in the coordinate frame),
 $\bF^i$ are the fluxes, and $\mathbf{S}$ is a vector of source terms. 
that do not involve derivatives of $\bP$ and therefore do not affect the
characteristic structure of the system.  
In non-relativistic MHD, both $\bP \rightarrow \bU$ and  $\bU
\rightarrow \bP$ have a closed-form solution. However, in GRMHD 
$\bU(\bP)$ is a complicated, nonlinear relation. Inversion
$\bP(\bU)$ is calculated 
once or twice in every time step, numerically.
The transformation between 'conserved' (momentum, energy density)
and 'primitive' (rest mass density, internal energy) variables requires to solve a set of 5 non-linear equations. 
This inversion is complex for a non-adiabatic relation of the pressure with density. We are doing it numerically and interpolate over the table of pre-computed EOS results.

\section{Visualization and Post-processing of Results}
\label{sec:kostas}

The code produces a set of outputs that can be divided into three categories. The Initialization Output is only produced once, before the integration begins, and contains the constant quantities of the simulation like the grid and
coordinates, the metric and its determinant on the grid points. Some model parameters are also stored during the Initialization stage, e.g. the BH spin parameter, the adiabatic index etc. 

The Results Output contains the main results of the integration and stores the physical quantities of the flow (density, internal energy, contravariant velocity, magnetic field vector), while it contains also some other derived quantities of physical interest. Among the various options we used for the form of the dumping data (text, raw binary, HDF5, etc), the HDF5 proved to be the most advantageous.
The 1-D, and 2-D models of an ideal conducting fluid does not pose significant restrictions on the dumped data type. The situation alters when we assume a composition dependent EOS or we proceed to the 3-D simulations where both file size and structure complexity increase dramatically. The hierarchical structure of the file makes easy
to locate specific quantities through a \textit{POSIX}-like syntax.

The size of the dumped file or of the objects it includes is not limited, while the special extensions, \textit{slib, zlib}, can be used to compress the resulting file. But the most significant features that HDF5 provided, 
were the performance of its collective (parallel) I/O driver and the portability of the data in both C/C++ and Python/IPython interfaces. The robust performance of the I/O driver was noticed in both of the HPC and local servers we used.

\begin{figure} 
 	\centering
 	\includegraphics [width=0.5\textwidth]  {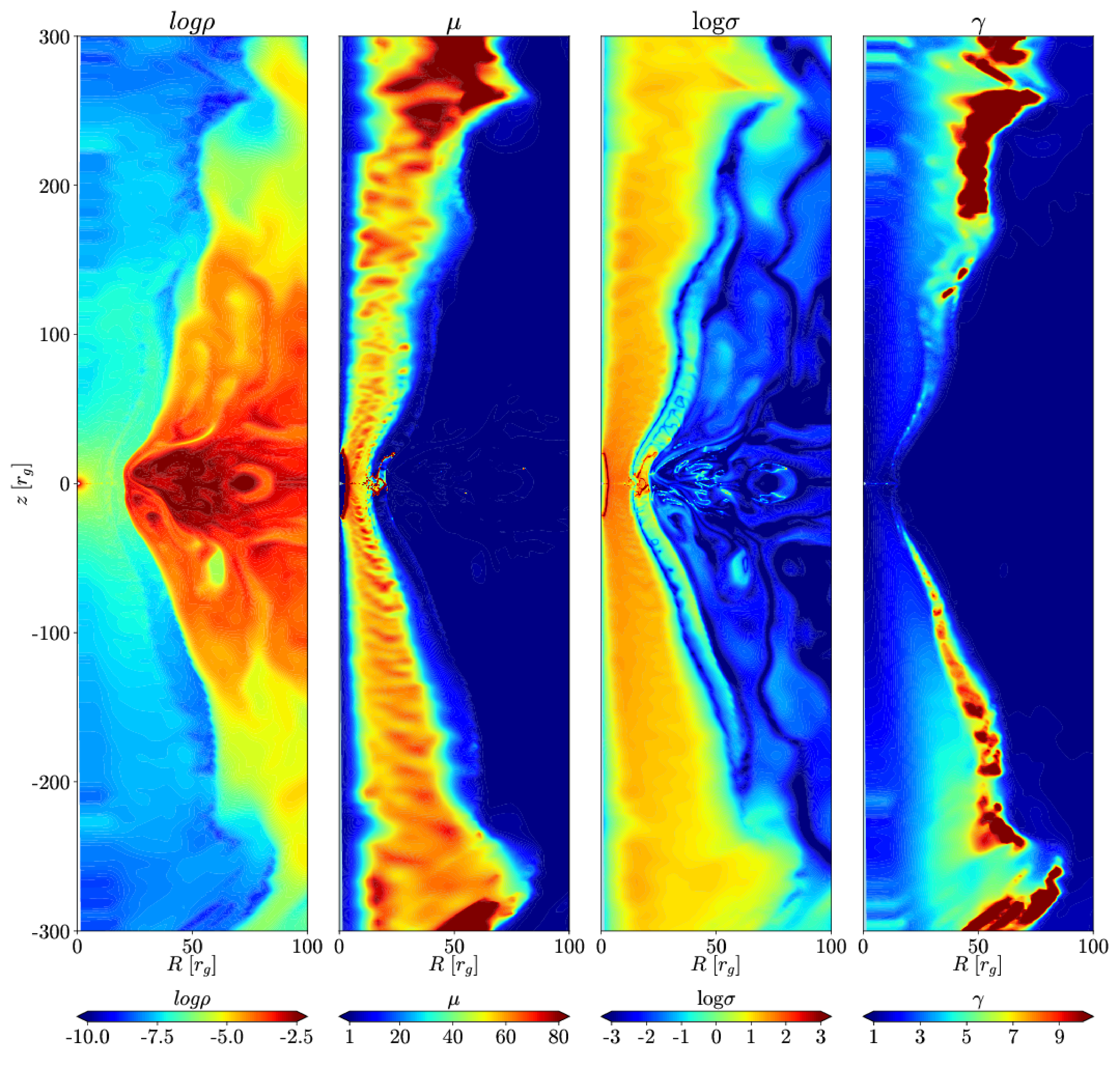}
 	\caption{Space distribution of the density, magnetization, and Lorentz factor in the jet-outflow in the GRB. Example of the graphical visualization with \textit{matplotlib/pyplot}}
\label{Jets}
\end{figure}

\begin{figure} 
\centering
\includegraphics[width=0.6\textwidth]{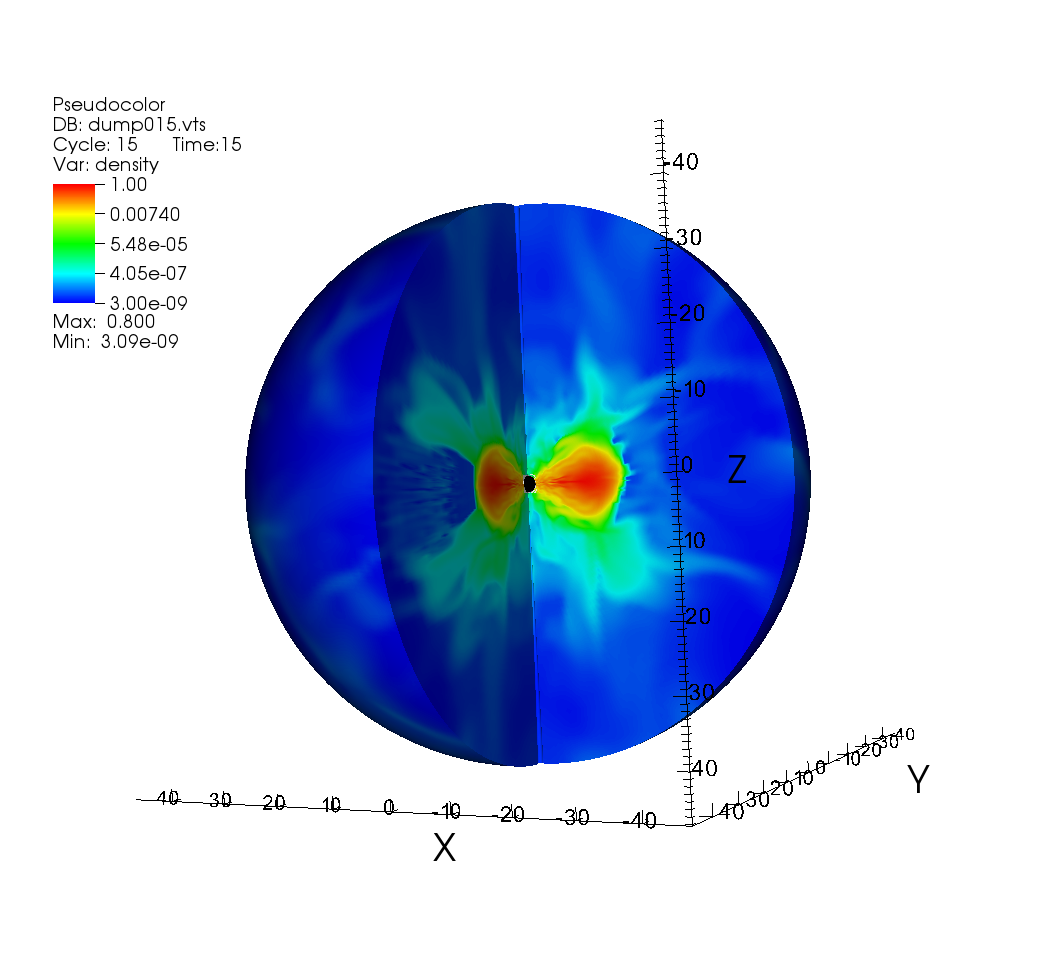}
\caption{3D simulation of the black hole-torus accretion system. The snapshot is taken at $t=1000 ~M$. The $x$, $y$ and $z$ axes are in $r_g$ units (gravitational radius). The black hole is represented with a black circle. Magnetic field is neglected in this simulation. The color scale shows the normalized density}
\label{3D}
\end{figure}

The post-processing and the visualization of the results for the 1-D, 2-D simulation is performed by the standard packages of the \textit{Python3} language (\textit{numpy, matplotlib/pylab, scipy}). With the extension of the \textit{h5py} package, the import of the results is straightforward and sequentially the powerful routines of the python libraries are used for further manipulation. An exemplary plot (Fig. \ref{Jets}) was produced using the \textit{matplotlib/pyplot} module. The parallelization of the above scripts proved to be a challenging task and we concluded that the \textit{mpi4py} module is easier portable to a series of machines from our Desktops and 32-Intel-cores local server, up to the Cray XC40 of the Okeanos HPC, mostly because of the parallel \textit{HDF5/h5py} and the \textit{mpi4py} module compatibility. It is clear that the parallelization is essential especially for the 3D post-process where various manipulations, e.g. interpolation, coordinate transformations, have to be done on the points of extended grids.  Once these calculations are completed, we use \textit{PyVtk} module to produce a VTK file output and more advanced tools for the 3D visualizations, e.g. \textit{VisIt}. An example plot is shown in Fig. \ref{3D}.


The Debug Output is the final set of output produced during the run time. Beyond the simple execution messages and the validity of the $\vec{\nabla}\cdot \vec{B} = 0$ condition, the code performs a number of physical diagnostics during each step of integration and dumps the results through a series of binary and ppm graphic files (per process). The motivation for these tests is to help user to identify unphysical results and avoid time consuming calculations.

To sum up, the HDF5 format is characterized by:
\begin{itemize}
\item Robust and satisfactory performance of the MPI I/O process.
\item High portability in many interfaces (C/C++ and Python).
\item Easy to locate quantities through the POSIX type structure. 
\end{itemize}

Furthermore, our Python main processing (cython under development) allows for:
\begin{itemize}
\item Calculation of various physical quantities (h5py). 
\item Building of VTK Cartesian Structured Grid (PyVTK). 
\item Parallel Python (mpi4py).  
\item 2D slices and analysis (numpy, scipy, etc).
\end{itemize}

Finally, we take advantage of the open-source tool \textit{VisIt} for visualization and 3D post-processing.

\section{The Boundary Conditions}
\label{sec:axis}

The HARM code does not solve the equations of the GRMHD in the Boyer-Lindquist coordinates system, but
rather on a modified version of the so called Kerr-Schild coordinate system (KS). The KS system is not singular on the black hole horizon and therefore, the matter can accrete smoothly through this surface and the evolution of the flow can be followed, see for example \cite{font1998}. A further transformation applies on the radial component of the specific system using a logarithmic mapping \cite{gammie2003harm}. As a consequence, our points distribution is denser close to the horizon, when we assume an equally spaced grid on the r-direction.


The boundary conditions that apply on the radial direction are the free
inflow-outflow conditions, modified by a specific extrapolation schema that reduces the unphysical behavior.
This behavior is induced mostly by the variation of the metric between the normal and the ghost cells \cite{gammie2003harm} and the selected extrapolation was chosen such as to maximize the robustness of the code.
In reality, the radial boundary conditions have negligible effect on the physics of the problem under the proper choice of the grid. The inner boundary is located inside the horizon, while due to the logarithmic spatial scale of the grid, the outer boundary can be set far away from the domain of interest.



The 2-dimensional simulations are, by definition, axisymmetric, i.e.the derivatives of quantities in the $\phi$-direction are neglected; note however, that the vectors (velocity field, magnetic field) still have all the three components. In contrast for the 3-dimensional simulations the derivatives of the quantities are computed so the non-axisymmetric evolution is being followed properly. The periodic boundary condition is always used in the azimuthal direction and the flow quantities at $\phi_{0}+2\pi$ are the same as at $\phi_{0}$ describing the smooth continuation of the solution between the neighboring domains. 

\begin{figure} 
\centering
\includegraphics[width=0.4\textwidth]{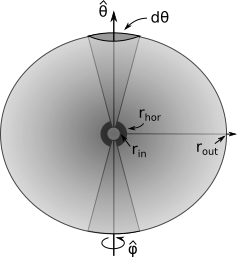}
\caption{Illustration of the spherical coordinate system and boundary conditions problem}
\label{axis}
\end{figure}

The real technical problem is posed by the boundary condition at the $z$-axis, namely in the polar direction.
Physically, the vertical axis is only the symmetry axis in the 2D flows, but it should not work as a real boundary in the 3D flows. In order to get some intuition on this effect, the reader can think of a Cartesian coordinate system. In such a case boundary conditions has to be set in an inner box, that will lay well inside the horizon and an outer one which lays at great distances, but not on the axis of the black hole rotation (see Fig. \ref{axis}). A technique to overcome the extra needed boundary conditions is to cancel in practice the axis existence by attaining the 'ghost cells' variables from the values of the corresponding normal grid cells on the other side of the axis; notice though that the vectors $\phi-$ components must change sign because of the opposite direction of the $\phi-$unitary. Therefore the application of the above technique requires by our algorithm to connect the two correspond grid domains and sets further complexity in our MPI implementation.

\section{Parallelization Methods in HARM Simulations}
\label{sec:mpi}

The HARM code works in 1, 2, and 3-Dimensions. In the latter case, the optimal 
time for any realistic simulation requires parallel computing.

In the simulation presented in Fig. (\ref{3D}), i.e., the non-magnetized, 3-dimensional case, 
the grid resolution was 192x192x192 points. The number of HPC nodes was 
N=64, the number of tasks per node was n=24. The number of run-time steps of 
integration was about 110000 and the total real time of computation was about 
12.7 hours.

\subsection{MPI in Practice}
\label{sec:general}

The distribution of the processes among the physical system directions is provided by the function \textit{MPI\_Dims\_create(nprocs, 2, dims)}; currently is 2D, but it can easily generalized to 3D. According to this routine the divisors are set to be as close as possible using an appropriate divisibility algorithm, while
\textit{dims[i]} are ordered in decreasing order:
$dims[0] \ge dims[1] \ge dims[2] \ge ...$. 

An alternative procedure incorporated also in our schema distributes the number of processes using two criteria. In order to reduce bottlenecks, the $N_{x} \times N_{y}$ grid is distributed to the $n_{x} \times n_{y}$ processes on each direction such as $N_{x}-n_{x}[{N_{x} \over n_{x}}]$ processes loaded with $[{N_{x} \over n_{x}}]$ points and $n_{x}([{N_{x} \over n_{x}}]+1)-N_{x}$ processes with $[{N_{x} \over n_{x}}]+1$ ones where the square brackets denote the integer part of the division; the same holds for the $y$ direction. We then perform an optimization routine requiring that the grid points load per each process is as balanced as possible, while in addition the necessary MPI communications between the neighboring domains are minimum. The two criteria are calculated with different weight allowing us to perform further optimization based on the specific integrating system and the machine specifications.

The assignment of the Cartesian grid to the MPI communicator is performed with the default MPI\_Cart\_create and MPI\_Cart\_coords routines that provide a specific rank per direction to every process. The latter is of primary importance for identifying the mirror to the rotation axis points and applying the proper boundary conditions (see Sec. \ref{sec:axis}).

\subsection{Supercomputer Performance Tests}
\label{sec:performance}
\begin{figure} 
 	\centering
\includegraphics [width=0.49\textwidth]  {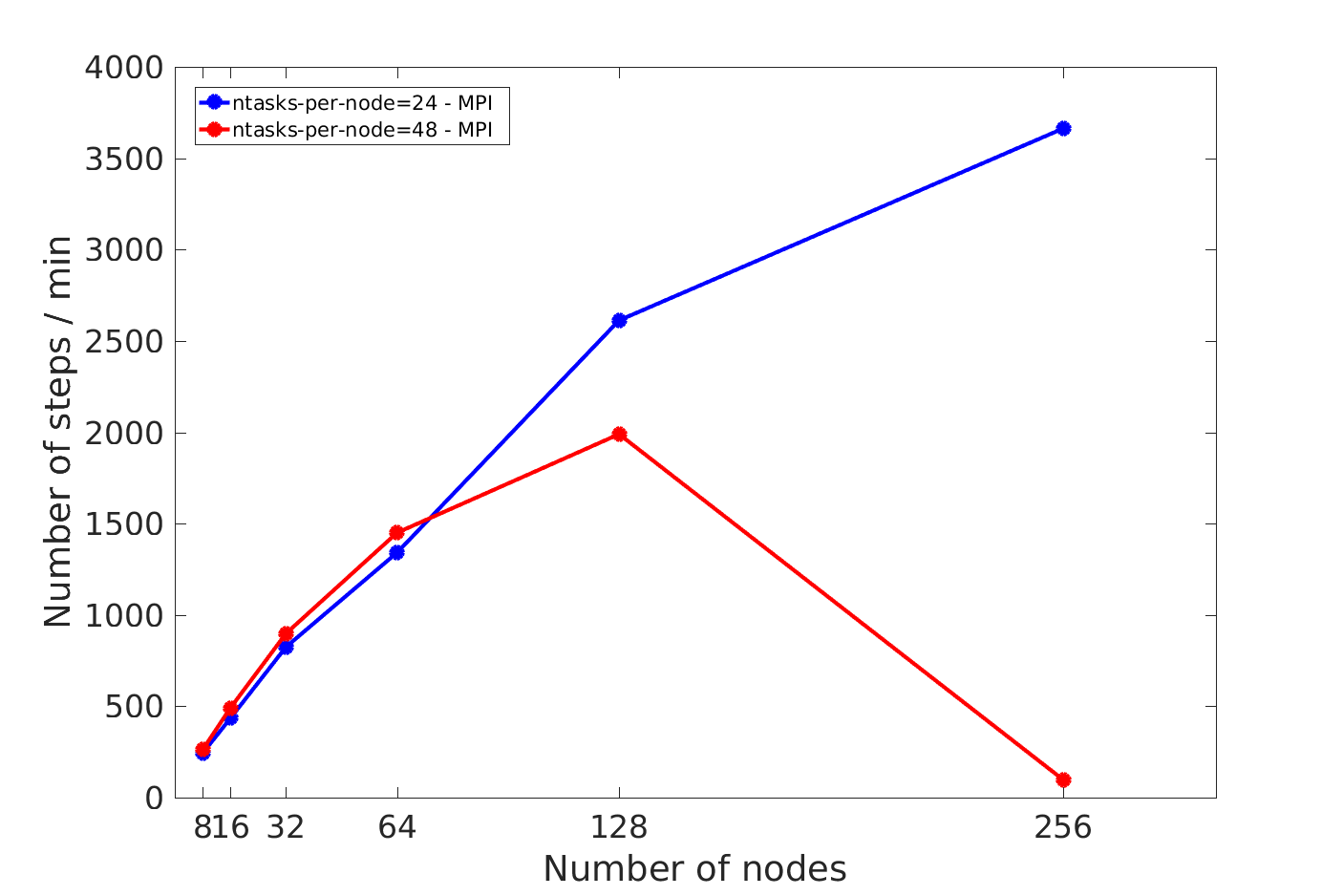}
\includegraphics [width=0.49\textwidth]  {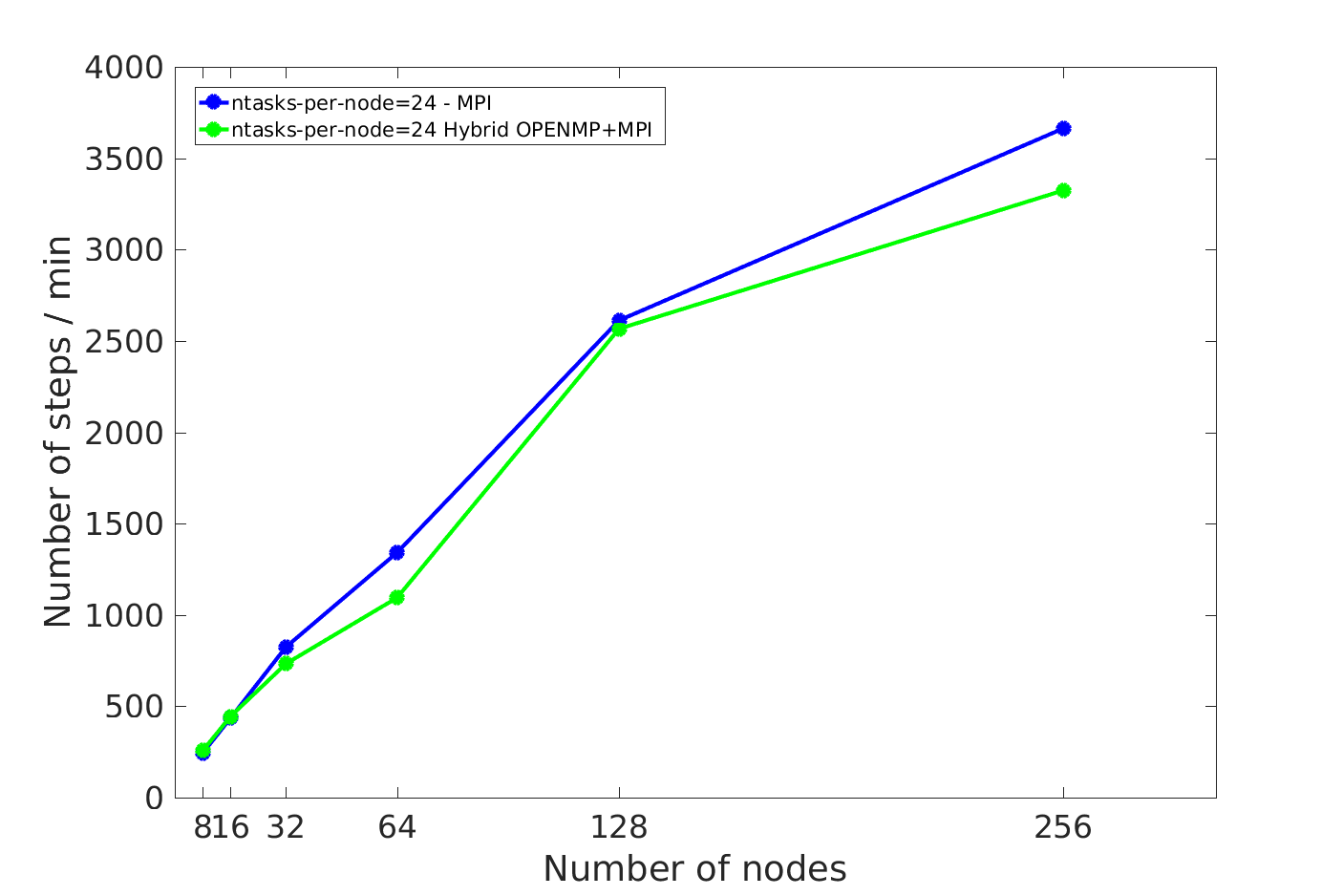}
 	\caption{Performance test results on the Okeanos supercomputer. The code HARM-2D was run on 8-256 nodes, using MPI parallelization method (blue line), and Hybrid MPI+OpenMPI method (green line). Also, the pure MPI runs were made with the hyper-threading (on 2x24=48 threads; shown by red line)}
\label{MPI}
\end{figure}

We used both the standard Message Passing Interface (MPI) technique, and also 
a more advanced, Hybrid parallelization method.
The MPI+hyper-threading was also tested
(cf. Fig. \ref{MPI}, red and blue lines).
The computational grid was divided into slices where every process works on its own area and boundaries are exchanged when needed. The pure MPI program running on the cluster of 1024 nodes and using 24 threads per node creates 24576 processes in total. It is a huge number, and the first expectation is that interprocess communication for boundaries exchange should decrease the overall performance.

The code occurred to be well-scalable for a uniform resolution, e.g. the number of the grid points in 3 dimensions equal to $N_{\rm r} \times N_{\theta} \times N_{\phi}=192 \times 192 \times 192$. For non-uniform grids, the dependence on $N_{\rm nodes}$ and $N_{\rm tasks/node}$ is not monotonic,
due to the properties of function \textit{MPI\_Dims\_create}.
Another method was to use the MPI + OpenMP shared memory, i.e., a Hybrid approach \footnote{http://bisqwit.iki.fi/story/howto/openmp/ and https://computing.llnl.gov/tutorials/openMP/}.
Here, only one MPI process is created per node and called master.
The parallel execution is done for every loop in the code in \textit{fork-join} mode.
It was rather straightforward (in comparison to MPI) to add OpenMP calls, using only several \textit{pragma} statements.  
Our preliminary tests were aimed to check if pure MPI-HARM, running on $N$ nodes with 24 cores each will be more/less efficient with comparison to MPI+OpenMP hybrid solution running on the same number of nodes.
The results are in contradiction to the claims published in literature, which show that the hybrid solution usually works better.

In our code, there is one significant difference with respect to the common MPI usage. We have computational domain divided into pieces and every MPI process uses only small memory region. This gives a faster memory access and it might make a difference.

\section{Models Specific to Black Hole Accretion Systems}
\label{sec:simulations}



Below, we present some exemplary results of our simulations.
Then, in the next Section, we discuss their results in the context of the visualization and post-processing methods. These 'technical' aspects are by no means trivial from the computational point of view, and proper analysis of the physics involved is tightly coupled with the post-processing demands.
Finally, these simulations are at the limits of our computational resources, and fine numerical techniques have to be used to increase the efficiency of the simulations and code performance on the available computer clusters.

\subsection{Magnetized Torus}

In the purpose of better understanding of the black hole accretion and
jets variability, we investigate the role of magnetically arrested disks (MADs) \cite{sasha_review}, as producer of turbulence in relativistic jet. MADs state occurs when the magnetic pressure force, pushes outward on the accretion disk gas.
Because we need a certain amount of magnetic flux surrounding the black hole, to have enough magnetic pressure balancing the accretion, we need a large initial magnetized torus. To do this, we implement a thick disk model as an intial condition in HARM using the Chakrabarti's prescription 
\cite{chakrabarti1985}. In this model, the angular momentum distribution is chosen to have a power law distribution.
Alternatively, the default disk model of Fishbone \& Moncrief \cite{Moncrief},
is assuming the angular momentum to be constant in the disk. This model allows to create an initial torus configuration with a large amount of poloidal magnetic flux.
\begin{figure} 
\centering
\includegraphics[width=0.8\textwidth]{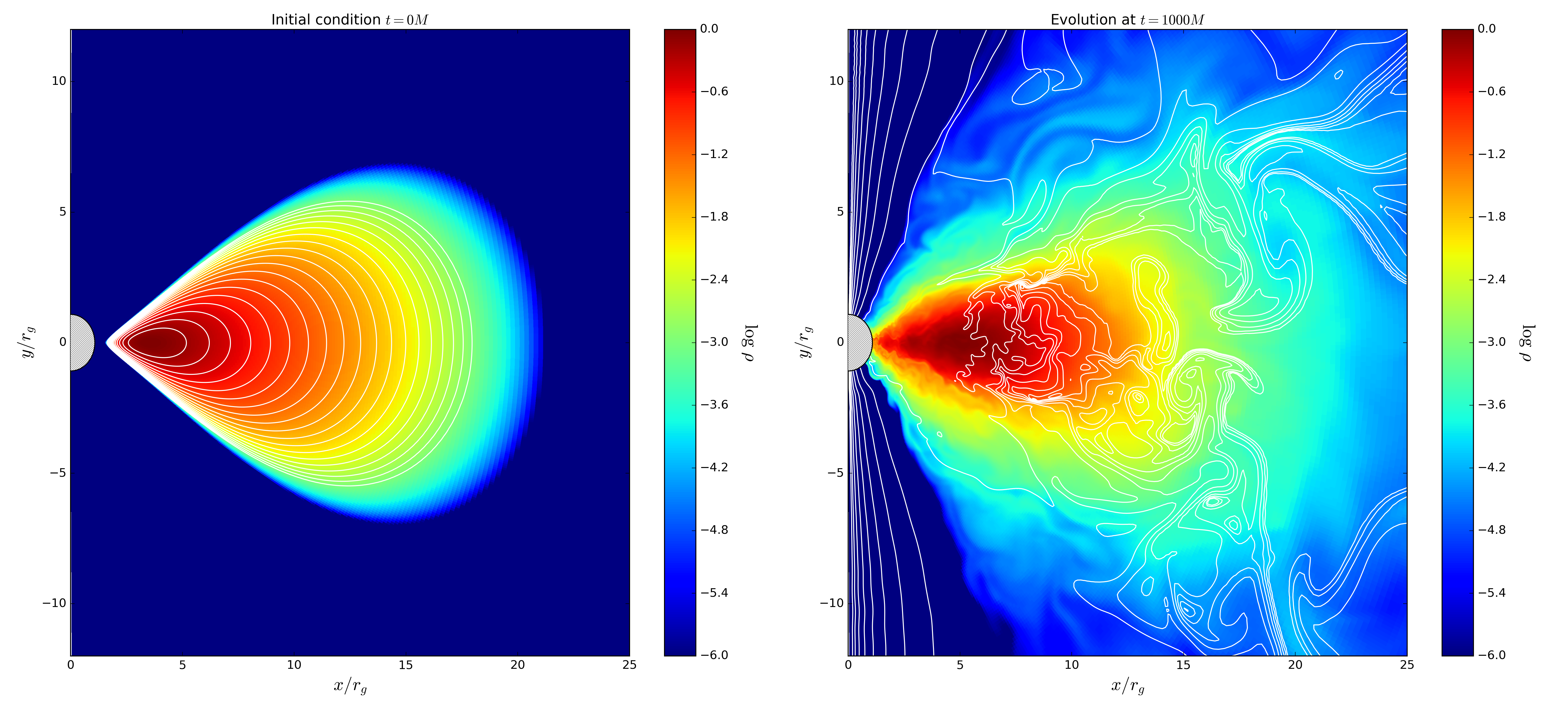}
\caption{2D simulation of a black hole-torus accretion system, with resolution of 512 x 512 grid points in the $r$ and $\theta$ direction. The left panel shows the initial condition, while the right one presents a snapshot a $t=1000 M$. The $x$ and $y$ axes are in $r_g$ units (gravitational radius). The black hole is represented with a dashed circle. Magnetic field lines are plotted in white contours, and the color scale shows the normalized density}
\label{Sim}
\end{figure}
The initial and evolved state of the trous, seeded by the poloidal magnetic field, is visualized in Fig. \ref{Sim}.

\subsection{Ejection of Relativistic Jets}

\begin{figure} 
 	\centering
\includegraphics[width=6.5cm]{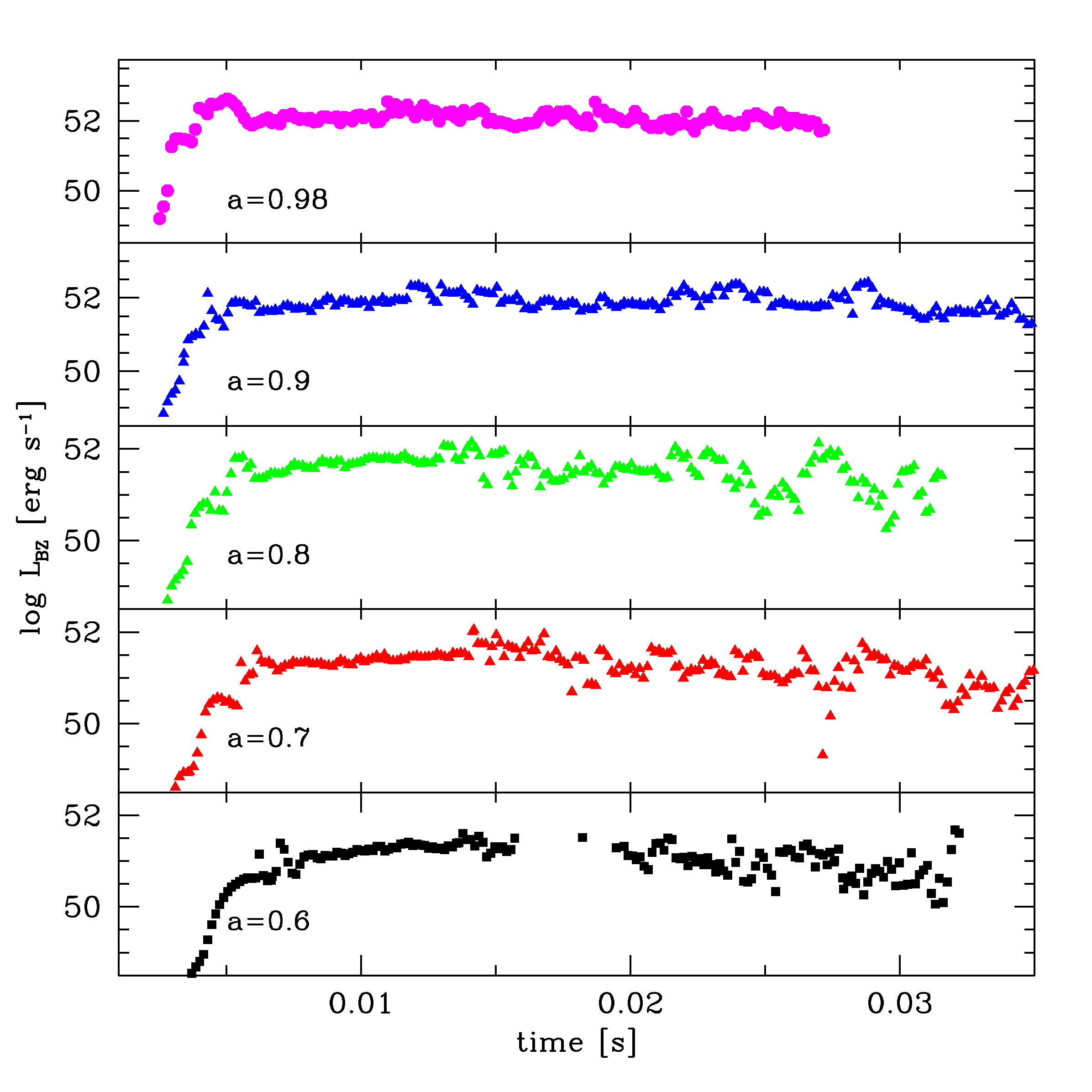}
\includegraphics[width=6.5cm]{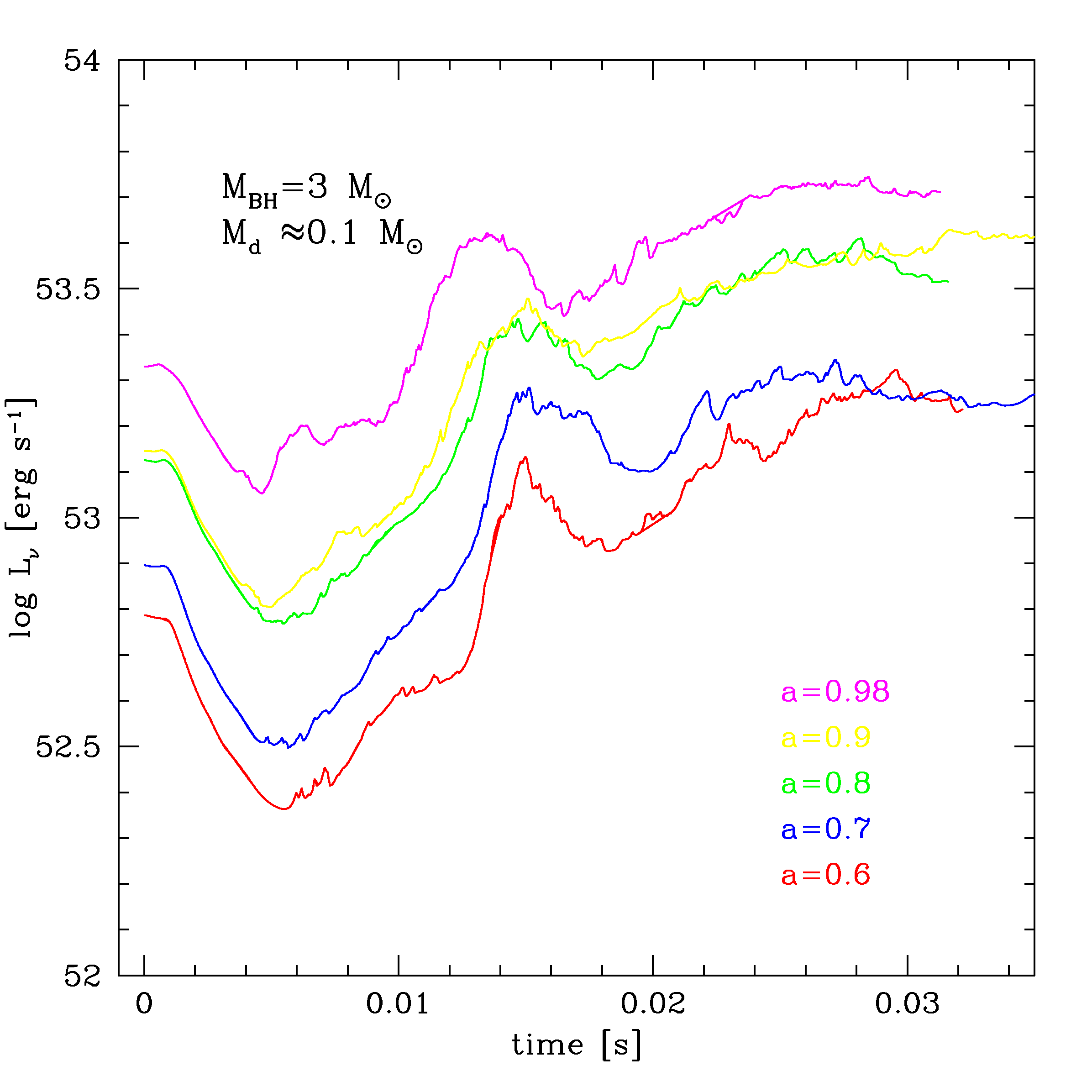}
\caption{Power available to accelerate the relativistic jets, produced within the accretion torus in the GRB central engine. Several results are plotted, for a varying black hole spin parameter, $a$, as denoted in the figures. Two mechanisms are compared: Blandford--Znajek process (left) and neutrino anihillation (right). Note, that the 
  latter has to be reduced by an efficiency factor, on the order of $\eta_{\nu}\sim0.01$, due to the uncertainties in the neutrino pair annihilation process. }
\label{fig:bzpower}
\end{figure}

If the black hole starts fastly rotate, the jet ejection is inevitable.
The presence of magnetic fields and/or neutrino-antintineutrino pairs
provide the mechanism for jets acceleration. 
The Blandford--Znajek process, which allows for the extraction of rotational energy of the black hole and transports it to the remote jets via magnetic fields,
 can be quantified in our simulations with the following expression
\begin{equation}
\dot{E} \equiv \int d\theta d\phi\, \sqrt{-g}\, {T^{r}}_{t} .
\label{eq:edotbz}
\end{equation}
Here, in Eq. (\ref{eq:edotbz}) the magnetic part of the stress-energy tensor is integrated over the black hole horizon. In addition, the magnetic fields transport angular momentum 
in the accretion disk and allow accretion.
In Fig. \ref{fig:bzpower} we show the resulting power available for the jets, as dependent on the black hole rotation spin.

The jets are accelerated at the vicinity of the black hole due to the central engine activity, and at large distances their kinetic energy has to be ultimately transferred to the emitted gamma ray radiation.
To achieve this, the jet speed, expressed with the dimensionless Lorentz factor,
\begin{equation}
\gamma= {1\over \sqrt {1 -({v \over c})^{2}}},
\nonumber
\end{equation}
has to be on the order of 100. This is required to avoid so-called 
'compactness problem', since the observed gamma ray spectra in GRBs are non-thermal. If the Lorentz factors in jets were small, the huge optical depth due to electron-positron pair creation would produce rather thermal emission
\cite{paczynski1986}.
The Lorentz factors of the jet estimated at 'infinity' in our simulations can easily reach the values around 80-100 (see Fig. \ref{Jets}).

\subsection{Formation of Heavy Nuclei in the Neutron-rich Plasma}

In the hyper-accreting disk at the GRB central engine, the conditions in the 
degenerated Fermi gas allow for substantial overabundance of the neutrons over protons.
This is quantified with so-called 'electron fraction' ratio:
\begin{equation}
Y_{e}={1 \over {1+n_{n}/n_{p}}},
\nonumber
\end{equation}
which in the highly neutronized matter is much smaller than 0.5.

The electron fraction distribution, together with the density and temperature, serve as an input for the subsequent nuclear reaction network computations \cite{Wallerstein, webnucleo, libnucnet, ajaniuk2017}.
The network allows for production of heavier isotopes (beyond Helium, and 
in fact beyond the Iron peak), due to the rapid 
capture of neutrons on the nuclei. The nuclear reactions may proceed with 1 (decays, electron-positron capture, photodissociacion), 2 (encounters), or 3 (triple alpha reactions) nuclei.
Abundances of the isotopes are calculated under the assumption of nucleon number and charge conservation for a given density, temperature and electron fraction ($T \le 1 MeV$).

In the GRB engine, along with the abundant light elements such as Carbon, 
and then Silicon, Sulfur and Calcium isotopes, we also
found copious amounts of Titanium, Iron, and Nickel.
The X-ray emission originating from the radioactive decay of 
isotopes, such as $^{45}Ti$, $^{57}Co$, $^{58}Cu$, $^{62}Zn$, $^{65}Ga$, $^{60}Zn$, $^{49}Cr$, $^{65}Co$, $^{61}Co$, $^{61}Cu$, and $^{44}Ti$, 
might give the signal in the 12-80 keV energy band.

The r-process elements have been found to enrich the interstellar gas, first
in our Solar system, and recently in the circum-burst environment of several Gamma Ray Bursts \cite{tanaka2016kilonova, tanvir2013kilonova} 
As discussed now in the literature, the dynamical ejecta launched during the  
compact binary mergers may be responsible for the of r-process nuclei.
In addition, the ejecta subsequently produced by an accretion disk formed after the merger, may add a contribution to this 'kilonova' lightcurve \cite{li1998transient}. 
Thus observational verification of our computations results will now be 
much more robust, thanks to the new data. The observed effect is discussed briefly below in Sec. \ref{sec:summary}.

\section{Summary}
\label{sec:summary}
 
Numerical modeling of black hole accretion flows in extremely high energetic
systems,
such as the gamma ray bursts, is essential from the point of view of the recent
observational discoveries. The discovery of gravitational waves in 2015, which was awarded the Nobel Prize in Physics in 2017, boosted the research also in high energy astrophysics, while it is related to the 'multi-messenger' astronomy. An example of the recently announced event is the kilonova signal, and the short gamma ray burst accompanied by the gravitational wave emission.

{\bf Kilonova effect}
\begin{itemize}
\item Optical and near-Infrared emission, powered by radioactive decay of r-process nuclei \cite{li1998transient}. 
\item Kilonova candidates can be distinguished from supernova by faster time evolution, fainter absolute magnitudes, and redder colors. 
\item Dynamical ejecta from compact binary mergers, $M_{\rm ej}\sim 0.01 M_{\odot}$, can emit about $10^{40}-10^{41}$ erg/s in a timescale of 1 week.
\item Subsequent accretion can provide bluer emission, if it is not absorbed by precedent ejecta \cite{tanaka2016kilonova}. 
\item In the GRB 130603B afterglow, the excess NIR flux corresponds to absolute magnitude $M(J)\sim 15.35$ mag at $\sim 7 $d after the burst, consistent with the kilonova behavior.
The lightcurve is in agreement with predicted r-process kilonova optical emission \cite{tanvir2013kilonova}. 
\end{itemize}

{\bf Electromagnetic counter part: GW 170816}
\begin{itemize}
\item Gravitational waves were discovered with the detection of binary black hole mergers and
they should also be detectable from lower mass neutron star mergers. 
\item NS-NS are predicted
to eject material rich in heavy radioactive isotopes that can power a kilonova.
\item  The gravitational wave source GW170817 arose from a binary neutron
star merger in the nearby Universe with a relatively well confined sky position and distance
estimate. 
\item A rapidly fading electromagnetic
transient in the galaxy NGC4993, is spatially coincident with GW170817
and a weak short gamma-ray burst \cite{smartt2017kilonova,cowperthwaite2017electromagnetic}. 
\end{itemize}

\section*{Conclusion}
\label{sec:concl}

With our simulations, we found that the proper microphysics treatment in GRMHD simulations of hyper-accretion is
essential for determining the disk structure: thickness, chemical composition
of torus and its ejecta. 
 Furthermore, we concluded that the neutrinos and Blandford--Znajek process have comparable role in powering the GRB jets.
As the large scale jets in GRBs are concerned, the variability of these jets is related to the disk's magneto-rotational turbulence timescale. The ultimate Lorentz factors are found to be on the order of few 100.
Thus, the late-time X-ray and high frequency radio emission can provide constraints on the properties of the disk-jet system for a particular 
source, e.g. GW+EM170817. 
Finally, the magnetically driven, low $Y_{\rm e}$ winds from accretion disks in GRB engine can provide power to kilonova emission, which was found in this source. 

\section*{Acknowledgments}
\label{sec:ack}
\ack{This work was supported by the grants 2012/05/E/ST9/03914, 2016/23/B/ST9/03114, and 2015/18/M/ST9/00541 from the Polish National Science Center. The computational resources were granted from the project GB 70-4 in the Interdisciplinary Center for Mathematical Modeling}
\openaccess

\bibliography{supercomp_revised2}


\end{document}